\documentclass[conference]{IEEEtran}
\IEEEoverridecommandlockouts
\usepackage[nocompress]{cite}
\usepackage{amsmath,amssymb,amsfonts}
\usepackage{algorithmic}
\usepackage{graphicx}
\usepackage[caption=false,font=footnotesize,labelfont=rm,textfont=rm]{subfig}
\usepackage{textcomp}
\usepackage{bm}
\usepackage{xcolor}
\usepackage{makecell}
\usepackage{multirow}
\def\BibTeX{{\rm B\kern-.05em{\sc i\kern-.025em b}\kern-.08em
    T\kern-.1667em\lower.7ex\hbox{E}\kern-.125emX}}
\begin{document}

\title{POST-MIMO: Partition-Oriented Semantic Transmission for MIMO Point Cloud Delivery
\thanks{This work was supported in part by Low-Altitude Airspace Strategic Program Portfolio under Grant Z25306105.}
}

\author{
	\IEEEauthorblockN{Hang Yue$^{1}$, Junlin Chang$^{3,4}$, Tianqing Zhao$^{3,4}$, Yulin Zhao$^{3}$, Rongke Liu$^{*,2,3}$\thanks{$^{*}$Corresponding author.}}
	\IEEEauthorblockA
	{
	\textit{$^{1}$National College for Excellent Engineers, Beihang University, Beijing, China} \\
	\textit{$^{2}$Shenzhen Institute, Beihang University, Shenzhen, China} \\
	\textit{$^{3}$School of Electronic and Information Engineering, Beihang University, Beijing, China} \\
	\textit{$^{4}$Department of Broadband Communication, Peng Cheng Laboratory, Shenzhen, China} \\
	\{yue\_hang, changjunlin, zhaotq, ylzhao, rongke\_liu\}@buaa.edu.cn
	}
}

\maketitle

\begin{abstract}
Point clouds have become a fundamental data format for emerging three dimensional (3D) applications, while their wireless delivery remains challenging due to the large data volume and irregular geometric structure. 
Existing semantic point cloud transmission methods are mainly developed for single-input single-output systems, making them difficult to apply directly to multiple-input multiple-output (MIMO) systems.
To address this issue, we propose a partition-oriented semantic transmission framework for MIMO (POST-MIMO) point cloud delivery. 
POST-MIMO partitions the input point cloud into spatially coherent blocks and aligns them with parallel MIMO streams. 
Based on this block-aligned representation, the transmitter employs a hierarchical encoder to capture both intra-block local geometry and inter-block global dependency, while the auxiliary side information is transmitted jointly to assist reconstruction. 
At the receiver, a progressive decoder reconstructs the block geometry in a progressive manner and restores the final point cloud with block-aware adjustment. 
Experiments demonstrate improved reconstruction quality and lower computational resource consumption than representative reference schemes under the considered MIMO setting.
\end{abstract}


\section{Introduction}
With the rapid development of three-dimensional (3D) sensing and modeling technologies, point clouds have become an important representation, whose effective representation requires preserving both fine-grained local structures and global context~\cite{chang2026point}.
Their large volume and irregular, unordered, and non-uniform structures, however, make efficient wireless delivery challenging~\cite{PCL}.

Conventional point cloud delivery follows the Shannon separation paradigm~\cite{Shannon}, with geometry compression followed by channel coding and digital modulation~\cite{GPCC}.
Although effective over reliable links, such bitstream-based transmission may degrade abruptly at low signal-to-noise ratios (SNRs).
Semantic communication provides an alternative by transmitting task-relevant or structurally meaningful information~\cite{shao2024,liang2026}.
In particular, Deep Joint Source-Channel Coding (DeepJSCC)~\cite{DeepJSCC} directly maps source data to channel symbols and has shown promise for point cloud transmission~\cite{sem_comm,OTA,Ying2025pc}.

Nevertheless, existing semantic point cloud transmission methods are predominantly designed for single-input single-output (SISO) systems and generally encode the entire point cloud into a globally coupled latent representation.
A straightforward multiple-input multiple-output (MIMO) adaptation splits this latent representation across parallel streams, but it does not explicitly establish correspondence between the spatial geometry and individual MIMO streams.
Therefore, a key research gap lies in developing a MIMO-oriented representation that maps spatially coherent geometric components to parallel streams while preserving both intra-component details and inter-component dependencies.

To bridge this gap, we propose a Partition-Oriented Semantic Transmission framework for MIMO (POST-MIMO) point cloud delivery.
POST-MIMO aligns spatially coherent blocks with parallel streams and employs hierarchical encoding to jointly model intra-block local geometry and inter-block global dependencies.
At the receiver, recovered auxiliary side information guides progressive geometric reconstruction and block adjustment.
Experiments demonstrate improved reconstruction quality and lower computational resource consumption than representative reference schemes under the considered MIMO setting.

\section{System Model}
\label{sec:system}
We consider a MIMO semantic communication system, where the transmitter with $N_t$ antennas delivers a point cloud to a receiver with $N_r$ antennas.
In this work, we focus on the spatial multiplexing setting with $N_r\ge N_t$.
The input point cloud is represented by the coordinate matrix $\bm{X}\in\mathbb{R}^{N\times 3}$, where $N$ is the number of points.
\begin{figure*}[!t]
    \centering
    \includegraphics[width=0.9\linewidth]{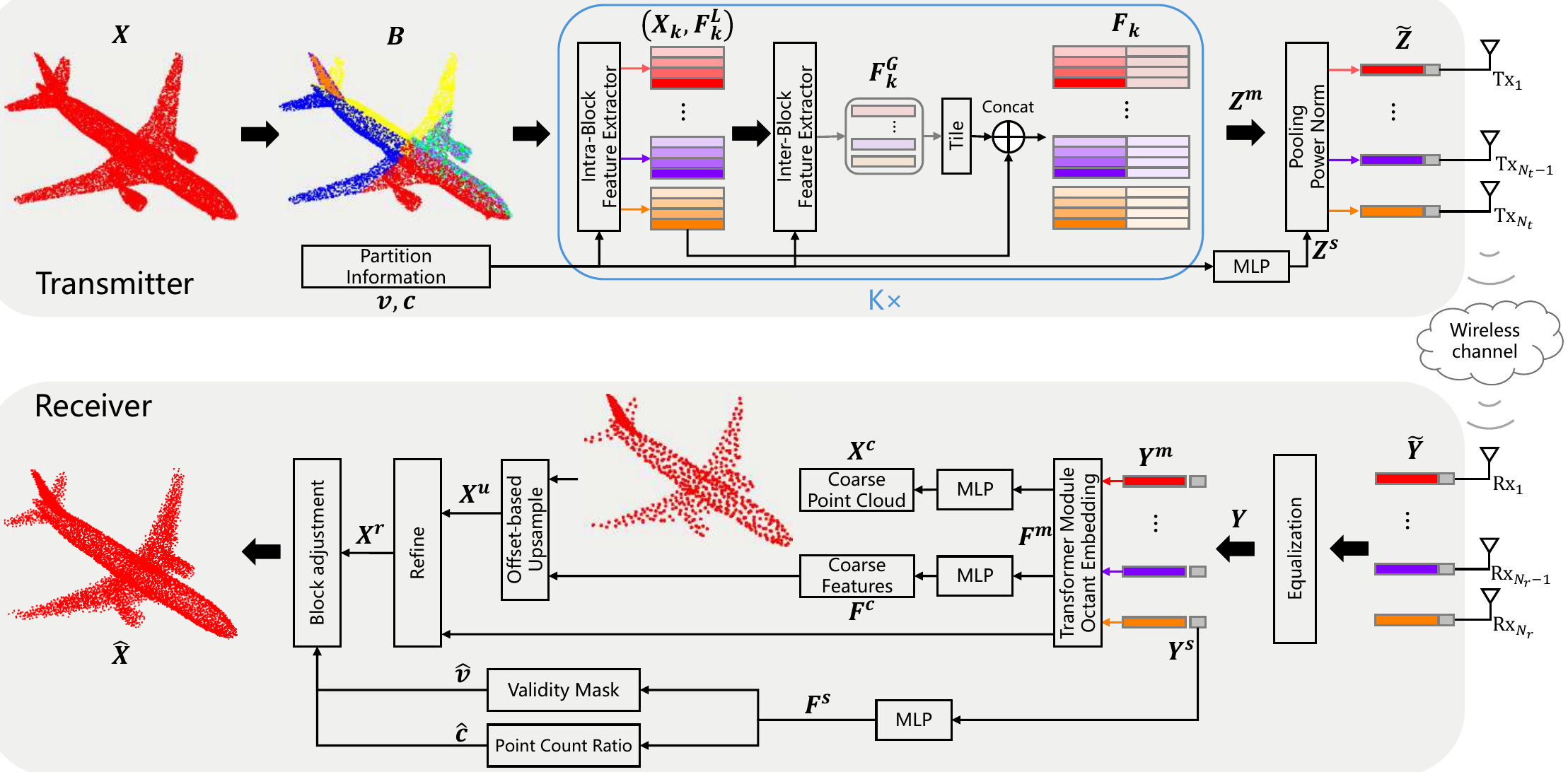}
	\caption{Overall pipeline of POST-MIMO.
	The transmitter partitions the input point cloud into spatial blocks, extracts the main feature and auxiliary side information, and maps them to MIMO channel symbols.
	At the receiver, the equalized streams are decoded through progressive geometric reconstruction and block adjustment to obtain the final reconstructed point cloud.}
	\label{fig:framework}
\end{figure*}
To exploit MIMO parallelism, the source $\bm{X}$ is partitioned into spatial blocks, denoted as $\mathcal{B}$.
Because the point density is spatially non-uniform, the number of points in each block may vary, and some blocks can be empty.
Therefore, two auxiliary metadata vectors are transmitted jointly: a validity mask $\bm{v}$ that indicates whether each block is empty, and a point count ratio $\bm{c}$ that records the point proportion of each block.

The transmitter contains two encoding branches, which produce the main feature $\bm{Z}^{m} \in \mathbb{R}^{N_t \times 2n_m}$ and the side information feature $\bm{Z}^{s} \in \mathbb{R}^{N_t \times 2n_s}$:
\begin{equation}
    \bm{Z}^{m}=f_{\mathrm{enc}}^{m}(\mathcal{B},\bm{v},\bm{c};\phi^{m}), \quad
    \bm{Z}^{s}=f_{\mathrm{enc}}^{s}(\bm{v},\bm{c};\phi^{s}),
\end{equation}
where $f_{\mathrm{enc}}^{m}(\cdot;\phi^{m})$ and $f_{\mathrm{enc}}^{s}(\cdot;\phi^{s})$ denote the main semantic encoder and the auxiliary side information encoder, respectively.
Both are implemented by neural networks with learnable parameters $\phi^{m}$ and $\phi^{s}$.
The two features are concatenated to form the real-valued codeword $\bm{Z}=[\bm{Z}^{m},\bm{Z}^{s}]\in\mathbb{R}^{N_t\times 2n}$, where $n=n_m+n_s$ denotes the total complex channel uses jointly allocated to the main semantic and auxiliary side information features.
The codeword is then converted to complex form $\tilde{\bm{Z}}\in\mathbb{C}^{N_t\times n}$ and normalized to meet the average transmit power constraint
\begin{equation}
    \frac{1}{nN_t}\|\tilde{\bm{Z}}\|_F^2 \le P_t,
    \label{eq:power_constraint}
\end{equation}
where $P_t$ denotes the average transmit power per antenna per complex symbol.

The received signal is
\begin{equation}
    \tilde{\bm{Y}}
    =\bm{H}\tilde{\bm{Z}}+\bm{W},
\end{equation}
where $\bm{H}\in\mathbb{C}^{N_r\times N_t}$ and $\bm{W}\in\mathbb{C}^{N_r\times n}$ have i.i.d. entries $H_{i,j}\sim\mathcal{CN}(0,\sigma_h^2)$ and $W_{i,j}\sim\mathcal{CN}(0,\sigma_w^2)$, respectively.
The channel is quasi-static, with $\bm{H}$ fixed within each transmission instance and independently resampled across instances.

At the receiver, perfect channel state information (CSI) is assumed for Minimum Mean Square Error (MMSE) equalization.
The equalized signal is then converted into the real-valued representation $\bm{Y} \in \mathbb{R}^{N_t\times 2n}$.
Finally, the decoder reconstructs the geometry:
\begin{equation}
    \hat{\bm{X}}=f_{\mathrm{dec}}(\bm{Y};\theta),
    \label{eq:decoder}
\end{equation}
where $\hat{\bm{X}}\in\mathbb{R}^{N\times 3}$ is the final reconstructed point cloud, and $\theta$ denotes the learnable parameters of the decoder.
Specifically, $\bm{Y}$ is split into $\bm{Y}^{m}$ and $\bm{Y}^{s}$: $\bm{Y}^{m}$ drives progressive geometric reconstruction, while $\bm{Y}^{s}$ guides block adjustment, including validity filtering and point count correction for the blocks.
The end-to-end learning objective is to reduce the geometric distortion between the input point cloud $\bm{X}$ and the reconstructed point cloud $\hat{\bm{X}}$.

\section{POST-MIMO Framework}
\label{sec:method}
This section presents the proposed POST-MIMO framework in detail. 
It begins with the block partition and stream alignment strategy for MIMO transmission, upon which the hierarchical semantic encoder and the progressive decoder are developed. 
In addition, the corresponding multi-task objective is introduced to jointly optimize geometric fidelity and auxiliary side information recovery.

\subsection{Spatial Partitioning Strategy}
\label{subsec:partition}
Given the input coordinates $\bm{X}\in\mathbb{R}^{N\times 3}$, we define the point set $\bm{X}=\{\bm{x}_1,\ldots,\bm{x}_N\}$ and partition it into $P$ octree blocks, denoted by $\{\mathcal{B}_p\}_{p=1}^{P}$, satisfying $\bigcup_{p=1}^{P}\mathcal{B}_p=\bm{X}$ and $\mathcal{B}_i\cap\mathcal{B}_j=\varnothing,\ \forall i\neq j$.
Here, $P$ denotes the total number of valid spatial blocks produced by octree partitioning, namely, the non-empty blocks that actually contain points.

Since one MIMO transmission instance can support at most $N_t$ independent spatial streams under the considered spatial multiplexing setting, the valid spatial blocks are scheduled group by group, with each group containing at most $N_t$ blocks.
When the total number of valid blocks exceeds the stream capacity of a single transmission instance, the blocks are divided into $T=\left\lceil P/N_t\right\rceil$ groups and delivered over $T$ successive transmission instances.
With a slight abuse of notation, after this temporal grouping we still use $P$ to denote the number of valid blocks in the current transmission instance for notational simplicity.
When $P<N_t$, the remaining stream slots are empty and will later be padded for tensor alignment. 

For each aligned transmission instance, we assign the $P$ valid blocks to the first $P$ streams and define the validity mask $\bm{v}\in\{0,1\}^{N_t}$ and point count ratio $\bm{c}\in[0,1]^{N_t}$ as
\begin{equation}
\begin{aligned}
v_i &=
\begin{cases}
1, & 1\le i\le P,\\
0, & P<i\le N_t,
\end{cases} \\
c_i &= v_i\frac{N_{b_i}}{N},
\qquad i=1,\ldots,N_t .
\end{aligned}
\label{eq:side_information}
\end{equation}
where $N_{b_i}=|\mathcal{B}_i|$ denotes the point count of the valid block.
The pair $(\bm{v},\bm{c})$ is transmitted as auxiliary side information for validity filtering and point-allocation recovery.

To enable parallel neural processing, the block size is aligned to the maximum point count among the blocks, i.e., $N_b=\max_{1\leq q\leq P} N_{b_q}$.
For each valid block, points are randomly resampled with replacement until the block contains $N_b$ points.
If $P<N_t$, the empty streams are filled with placeholders constructed by duplicating the densest valid block, which provides a spatial reference for the missing streams.
The aligned tensor is therefore written as $\tilde{\bm{\mathcal{B}}}=[\tilde{\mathcal{B}}_1,\ldots,\tilde{\mathcal{B}}_{N_t}] \in \mathbb{R}^{N_t\times N_b\times 3}$.
The placeholders are introduced only for tensor alignment, and their effects are explicitly suppressed by the validity mask $\bm{v}$ in downstream modules.

\subsection{Hierarchical Semantic Encoder Architecture}
\label{subsec:encoder}
\begin{figure}[!t]
	\centering
	
	\subfloat[]{
		\includegraphics[width=0.36\linewidth]{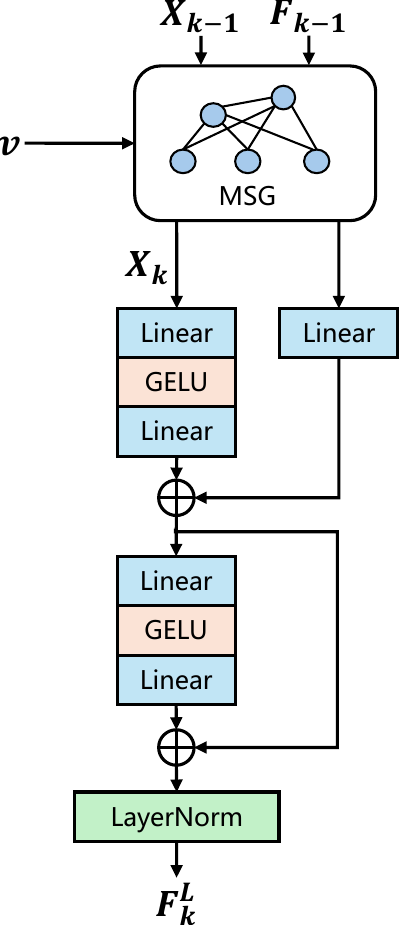}
		\label{fig:intra}
	}
	\hfill
	\subfloat[]{
		\includegraphics[width=0.44\linewidth]{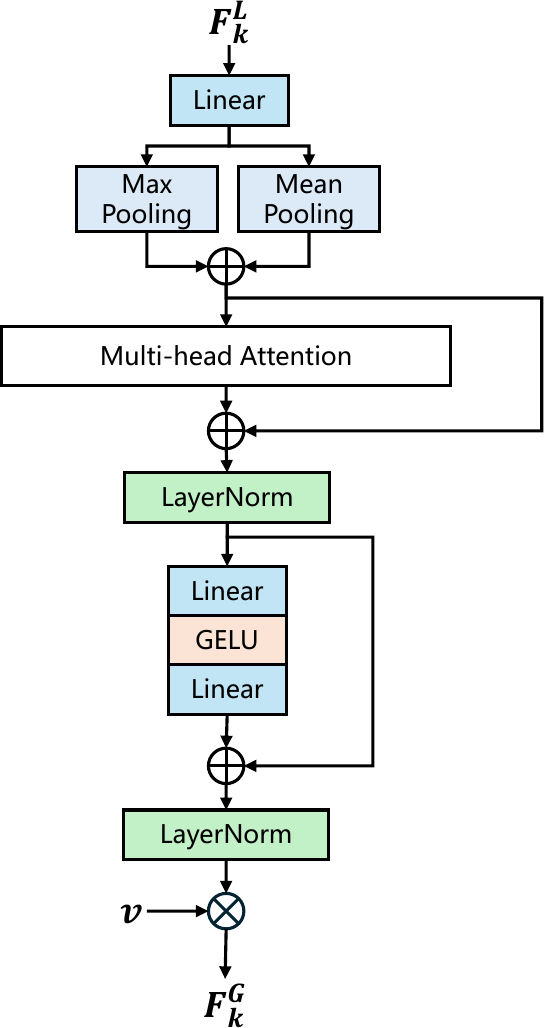}
		\label{fig:inter}
	}
	\caption{Architecture of the hierarchical encoder.
	(a) Intra-block local geometry modeling.
	(b) Inter-block global context aggregation.}
	\label{fig:encoder}
\end{figure}
The encoder is designed to preserve both local geometry within each block and structural dependency across different blocks.
To this end, it adopts a hierarchical architecture with $K$ stages of alternating intra-block and inter-block processing, followed by pooling to obtain the channel representation.

The hierarchical encoder operates stage by stage.
At stage $k\in\{1,\ldots,K\}$, the encoder takes the coordinates $\bm{X}_{k-1}\in\mathbb{R}^{N_t\times N_{k-1}\times 3}$ and the point-wise features $\bm{F}_{k-1}\in\mathbb{R}^{N_t\times N_{k-1}\times d_{k-1}}$ as input, where $N_k$ denotes the number of retained points per block after stage $k$, and $d_k$ denotes the corresponding feature dimension.
At the beginning of the hierarchy, we initialize both inputs from the padded block, i.e., $\bm{X}_0=\tilde{\bm{\mathcal{B}}}$ and $\bm{F}_0=\tilde{\bm{\mathcal{B}}}$.

As illustrated in Fig. \ref{fig:encoder}\subref{fig:intra}, the intra-block module extracts local geometric features independently within each valid block. 
This module is implemented by a PointNet++ \cite{pointnet2} set abstraction layer with multi-scale grouping (MSG), which performs local sampling and neighborhood aggregation while reducing the number of points. 
The output of the intra-block module is denoted by $\bm{X}_k$ and $\bm{F}_k^L$, where $\bm{X}_k\in\mathbb{R}^{N_t\times N_k\times 3}$ contains the sampled coordinates and $\bm{F}_k^L\in\mathbb{R}^{N_t\times N_k\times d_k^L}$ contains the local features. The validity mask $\bm{v}$ suppresses the contribution of placeholder blocks in this process.

Local modeling alone does not preserve the geometric dependency among spatially separated blocks. 
The inter-block module in Fig. \ref{fig:encoder}\subref{fig:inter} addresses this issue by constructing one token for each block and performing self-attention across all the streams. 
The block token $\bm{T}_{k}$ is formed by summing the max-pooled and mean-pooled features of $\bm{F}_{k}^{L}$ along the point dimension.
A standard Transformer~\cite{attention} block is then used to model inter-block dependency among these tokens.
The resulting block-level global feature is denoted by $\bm{F}_k^G\in\mathbb{R}^{N_t\times d_k^G}$, where invalid streams are suppressed by $\bm{v}$ before the global context is propagated to the next stage.

The global feature is broadcast back to the point level and fused with the local feature to form the stage output. Denoting the repeated version of $\bm{F}_k^G$ along the point dimension by $\bar{\bm{F}}_k^G\in\mathbb{R}^{N_t\times N_k\times d_k^G}$, the fused feature is written as
$
	\bm{F}_k
	=
	\mathrm{MLP}_k\left( \left[\bm{F}_k^L,\bar{\bm{F}}_k^G\right]
	\right),	
$
where $\mathrm{MLP}_k(\cdot)$ denotes the multilayer perceptron at stage $k$, which maps the concatenated local and global features to the stage output.
The pair $(\bm{X}_k,\bm{F}_k)$ is subsequently passed to the next stage of the hierarchy.
After $K$ stages, the encoder has access to the extracted coordinates $\bm{X}_K\in\mathbb{R}^{N_t\times N_K\times 3}$ and the corresponding features $\bm{F}_K\in\mathbb{R}^{N_t\times N_K\times d_K}$. 

In parallel, the auxiliary side information encoder processes the block metadata $(\bm{v},\bm{c})$ to generate the auxiliary feature $\bm{Z}^{s}$.
The main feature $\bm{Z}^{m}$ is then pooled and dimension-adjusted before being concatenated with $\bm{Z}^{s}$ for transmission.
This design allows the transmitted representation to preserve geometric features together with the block occupancy information required by the decoder.

\subsection{Progressive Geometric Reconstruction Decoder}
\label{subsec:decoder}
\begin{figure*}[!t]
	\centering
	\includegraphics[width=0.9\textwidth]{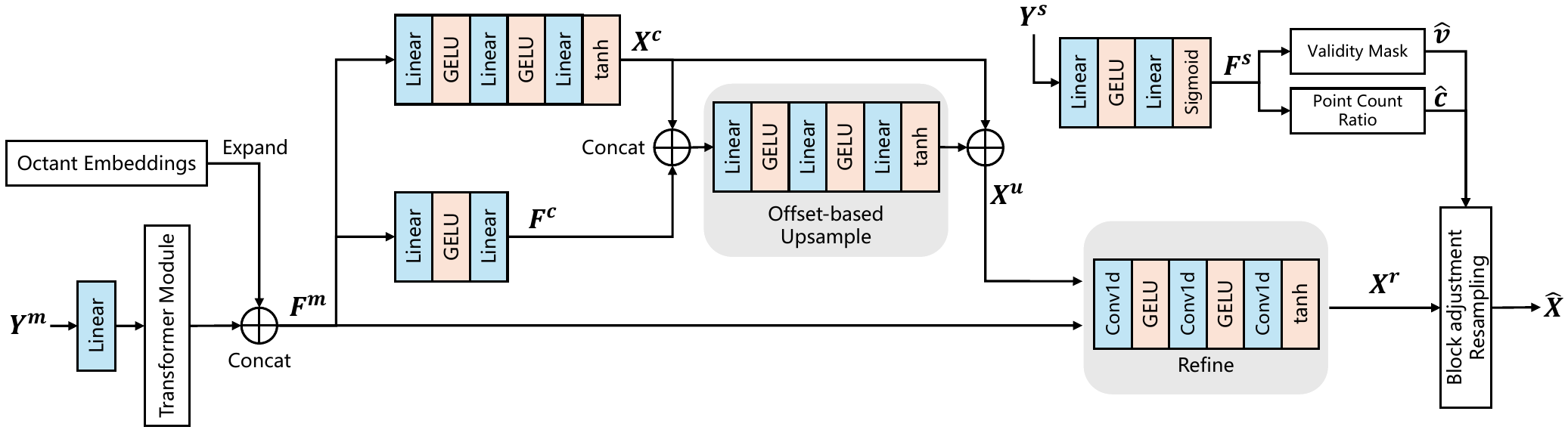}
	\caption{Architecture of the proposed decoder.
	The main signal and the auxiliary side information signal are jointly processed to reconstruct the point cloud in a progressive manner, followed by block adjustment and resampling.}
	\label{fig:decoder}
\end{figure*}
The equalized signal $\bm{Y}$ is split into the main signal $\bm{Y}^m$ and the auxiliary side information signal $\bm{Y}^s$.
The decoder uses $\bm{Y}^m$ to recover the geometry and $\bm{Y}^s$ to recover the occupancy and point allocation information, as illustrated in Fig. \ref{fig:decoder}.

The main branch starts from a linear projection of $\bm{Y}^m$, followed by a Transformer module that captures the dependency across the block streams.
To preserve the spatial identity of each block, learnable location embeddings are concatenated with the transformed block tokens, yielding the main representation $\bm{F}^m$.
In parallel, the side information branch maps $\bm{Y}^s$ to the estimated validity mask $\hat{\bm{v}}\in[0,1]^{N_t}$ and point count ratio $\hat{\bm{c}}\in[0,1]^{N_t}$ through a lightweight MLP.

Based on $\bm{F}^m$, two parallel heads generate the coarse block geometry and the corresponding coarse features, denoted by $\bm{X}^c\in\mathbb{R}^{N_t\times N_c\times 3}$ and $\bm{F}^c\in\mathbb{R}^{N_t\times N_c\times d_c}$, respectively.
Here, $N_c$ denotes the number of coarse points produced for each block and $d_c$ denotes the dimension of the coarse feature.

Dense reconstruction is obtained through an offset-based upsampling module.
For the $i$-th coarse point $\bm{x}_{p,i}^{c}$ in block $p$, an MLP predicts $r=N/(N_tN_c)$ local offsets from the concatenated coarse coordinate and feature pair.
The resulting upsampled points are given by
\begin{equation}
	\bm{x}_{p,i,j}^{u}=\bm{x}_{p,i}^{c}+\Delta\bm{x}_{p,i,j},
	\qquad
	j=1,\ldots,r,
\end{equation}
where $\Delta\bm{x}_{p,i,j}$ denotes the predicted offset.
By collecting all upsampled points generated from the $N_c$ coarse points in each block, we obtain the intermediate upsampled reconstruction $\bm{X}^{u}$.
A lightweight refinement network composed of cascaded convolutions further adjusts $\bm{X}^{u}$ together with the main-branch context, producing the refined reconstruction $\bm{X}^{r}$.

The final output is obtained by block adjustment and resampling guided by the decoded side information.
Invalid blocks are identified by hard thresholding $\hat{\bm{v}}$ and discarded, while points in each valid block are resampled according to $\hat{\bm{c}}$.
A lightweight correction compensates for rounding-induced integer discrepancies to ensure exactly $N$ points, thereby restoring the global point distribution and yielding the final reconstruction $\hat{\bm{X}}$.

\subsection{Multi-Task Optimization Objective}
\label{subsec:loss}
The network is trained end to end using a multi-task objective that jointly optimizes geometric reconstruction and side information prediction.

The geometric loss supervises both the coarse output and the final reconstructed point cloud, which is defined as
\begin{equation}
	\mathcal{L}_{\mathrm{geo}}
	=
	\mathrm{CD}(\bm{X}^{c}, \bm{X})
	+
	\mathrm{CD}(\hat{\bm{X}}, \bm{X}),
\end{equation}
where $\mathrm{CD}(\cdot,\cdot)$ is the Chamfer distance \cite{CD}.

Meanwhile, the side information branch is supervised using the aligned metadata of the $N_t$ transmission streams. 
Specifically, the validity mask is optimized by the binary cross-entropy loss:
\begin{equation}
	\mathcal{L}_{\mathrm{valid}}
	=
	-\frac{1}{N_t}
	\sum_{i=1}^{N_t}
	\left[
	v_i \log(\hat{v}_i)
	+
	(1-v_i)\log(1-\hat{v}_i)
	\right],
\end{equation}
while the point count ratio is optimized by the mean squared error
\begin{equation}
	\mathcal{L}_{\mathrm{count}}
	=
	\frac{1}{N_t}
	\sum_{i=1}^{N_t}
	(c_i-\hat{c}_i)^2.
\end{equation}

Thus, the total training objective is written as
\begin{equation}
	\mathcal{L}_{\mathrm{total}}
	=
	\lambda_{\mathrm{geo}}\mathcal{L}_{\mathrm{geo}}
	+
	\lambda_{\mathrm{side}}(\mathcal{L}_{\mathrm{valid}} + \mathcal{L}_{\mathrm{count}}),
\end{equation}
where $\lambda_{\mathrm{geo}}$ and $\lambda_{\mathrm{side}}$ balance geometric reconstruction and side information prediction, respectively.
Since the point coordinates are normalized during preprocessing, the geometric loss is numerically small.
In the experiments, we set $\lambda_{\mathrm{geo}}=1000.0$ and $\lambda_{\mathrm{side}}=10.0$ to maintain stable optimization.

\section{Experimental Results}
\label{sec:experiment}
We evaluate the proposed POST-MIMO framework on the ModelNet40 dataset \cite{modelnet40}.
Each point cloud is downsampled to 16,384 points using farthest point sampling and normalized to the range $\left[-1,1\right]$.
The network is trained using AdamW \cite{AdamW} with an initial learning rate of $5\times 10^{-5}$ and a cosine decay scheduler with 10 warm-up epochs.
The model is trained for 100 epochs with a batch size of 16.
All experiments are conducted on an Ubuntu 20.04 server equipped with an Intel Xeon Gold 5218 CPU and an NVIDIA A100 GPU with 40 GB memory.
Unless otherwise specified, we set $N_t=N_r=8$, $n=512$, and $n_s=16$ for the neural schemes in all experiments.
We evaluate reconstruction quality using D1-PSNR, D2-PSNR~\cite{PSNR}, and Chamfer distance (CD) \cite{CD}.
Higher D1-PSNR and D2-PSNR indicate better geometric fidelity, while lower CD indicates smaller geometric distortion.

We compare POST-MIMO with three representative baselines, namely G-PCC \cite{GPCC}, SEPT \cite{SEPT}, and DPCC \cite{DPCC}.
G-PCC is a conventional geometry-based point cloud compression standard.
SEPT is a DeepJSCC-based point cloud transmission framework that learns a global latent representation for geometry reconstruction.
DPCC is a learned point cloud compression method with neural encoding and decoding modules.
Since the baselines were originally designed for SISO transmission, we construct conservative MIMO adaptations by evenly splitting their transmitted latent vectors across the $N_t$ streams and concatenating the recovered streams before the original decoder.
These adaptations use the same channel-use budget as POST-MIMO for the neural schemes, but they do not include the proposed block-stream alignment.
For the digital baselines, the compressed bitstreams are additionally protected by a rate-$1/2$ Polar code before transmission.
Besides, DPCC is paired with BPSK modulation, while G-PCC is paired with 16QAM modulation.
It should be noted that G-PCC suffers from a severe bandwidth mismatch in the target low-bandwidth regime, and we set $n=5000$ in the experiments. Despite this generous bandwidth allocation, G-PCC still fails to achieve operable reconstruction.

\begin{figure}[!t]
	\centering
	\subfloat[]{
		\includegraphics[width=0.3\textwidth]{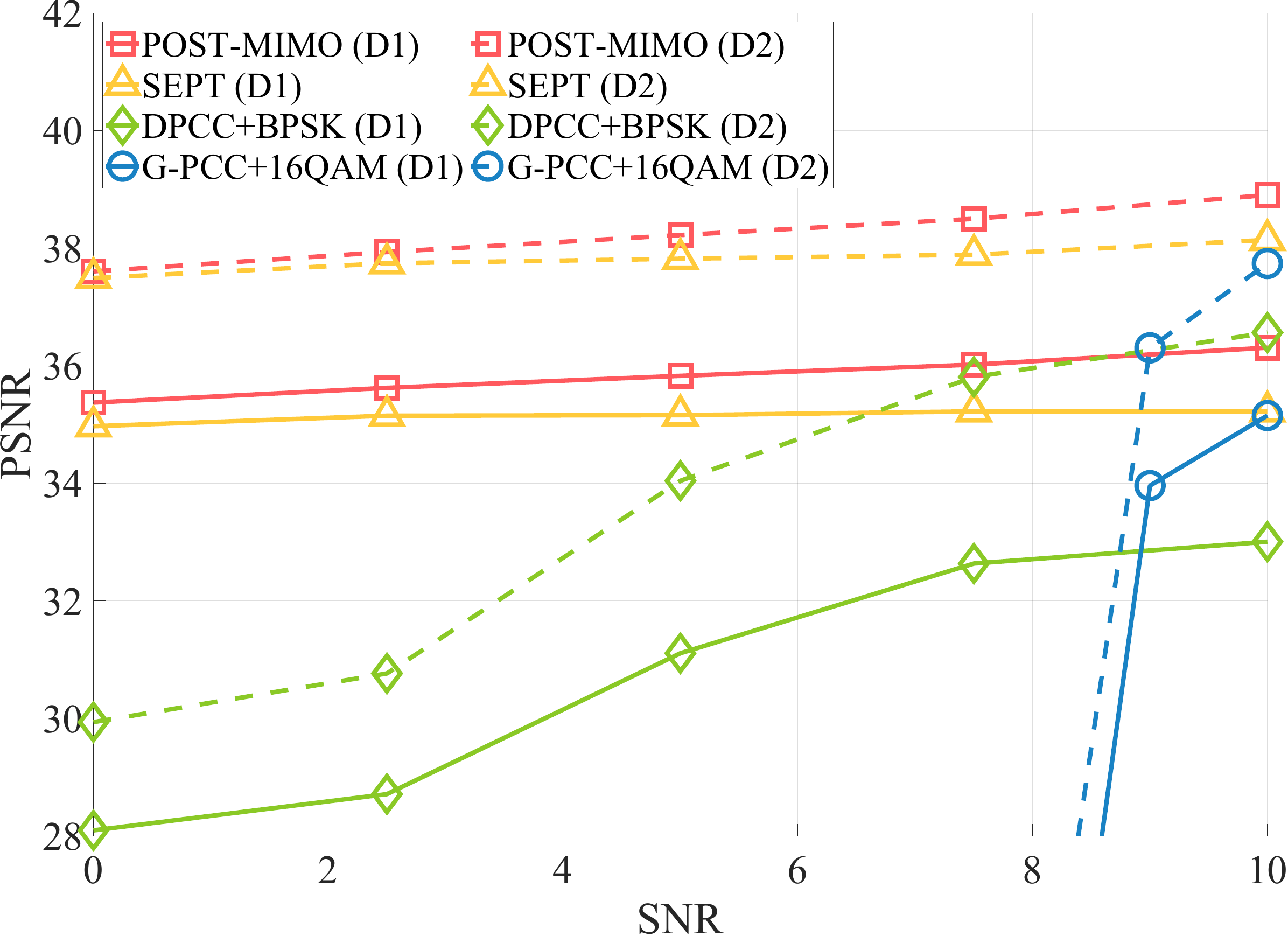}%
		\label{subfig:snr_psnr}
	}
	\hfil
	\subfloat[]{
		\includegraphics[width=0.3\textwidth]{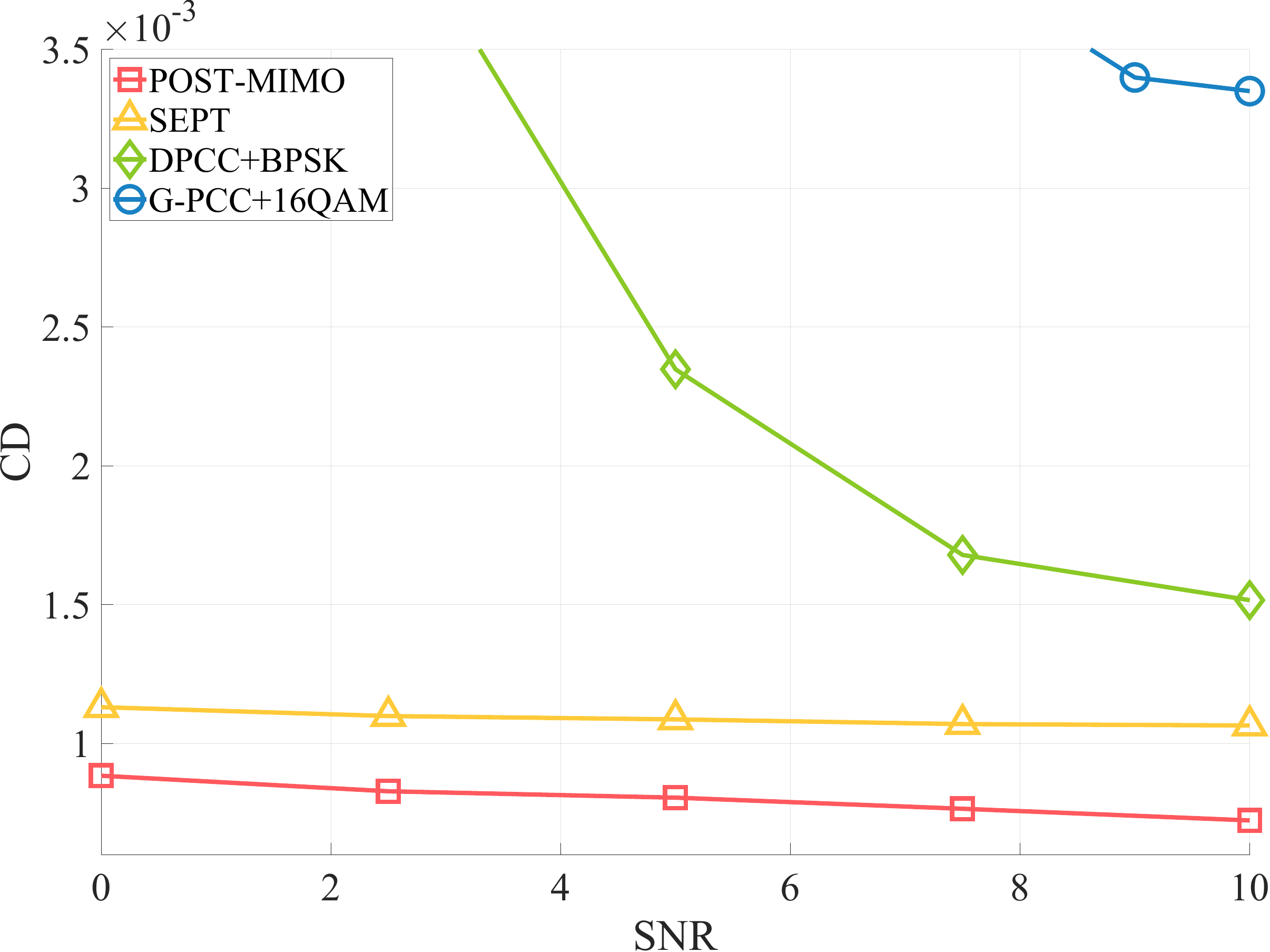}%
		\label{subfig:snr_cd}
	}
	\caption{Reconstruction performance under varying SNR.
	Here and hereafter, D1 and D2 denote D1-PSNR and D2-PSNR, respectively.
	(a) D1/D2.
	(b) CD.}
	\label{fig:snr}
\end{figure}

Fig. \ref{fig:snr} shows the reconstruction performance under different SNRs.
Under the current reference setting, G-PCC exhibits a clear threshold behavior and becomes ineffective once the SNR falls below a moderate level, which is consistent with the cliff effect of separate source and channel coding transmission.
DPCC can still operate in harsher channels, but its overall reconstruction quality remains inferior to other learned semantic schemes.
Both DeepJSCC-based schemes degrade more smoothly as the SNR decreases, while POST-MIMO achieves slightly better overall reconstruction quality than SEPT across the tested SNR range.
This gain is attributed to the block-aligned representation, which reduces the mismatch introduced by directly splitting a globally coupled latent representation across MIMO streams.
It is also worth noting that SEPT achieves competitive reconstruction quality under this conservative MIMO adaptation, but it requires substantially higher computational resources, as will be shown later.

\begin{figure}[!t]
	\centering
	% 定义统一的图片宽度，5列约 0.19 倍的行宽
	\newcommand{\figw}{0.18\linewidth}

	% 减小列间距，让图片尽量占满版面
	\setlength{\tabcolsep}{1.5pt} 
	
	\begin{tabular}{ccccc}
		% ================== 表头 ==================
		\makecell{\scriptsize Ground \\ \scriptsize Truth} & 
		\scriptsize SEPT & 
		\makecell{\scriptsize DPCC \\ \scriptsize +BPSK} & 
		\makecell{\scriptsize G-PCC \\ \scriptsize +16QAM} & 
		\scriptsize POST-MIMO \\
		
		% ================== 1. airplane ==================
		\includegraphics[width=\figw]{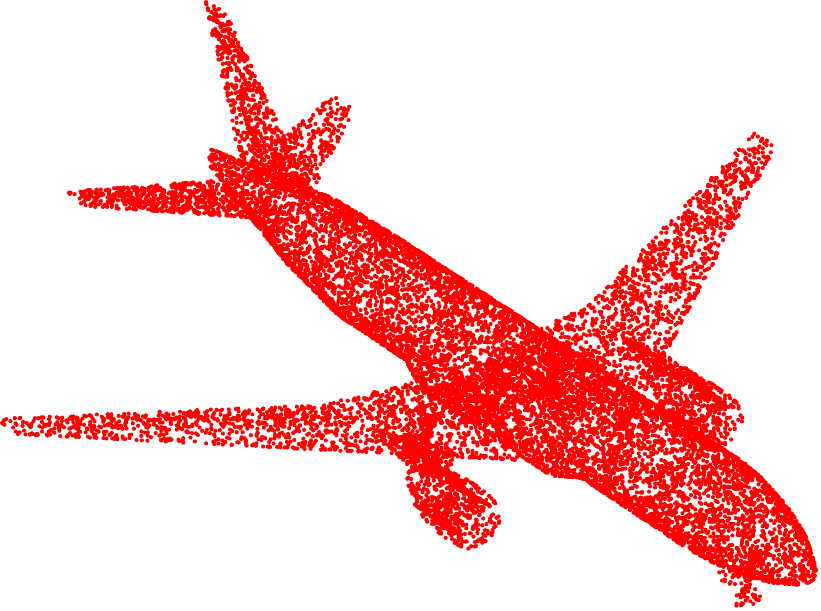} &
		\includegraphics[width=\figw]{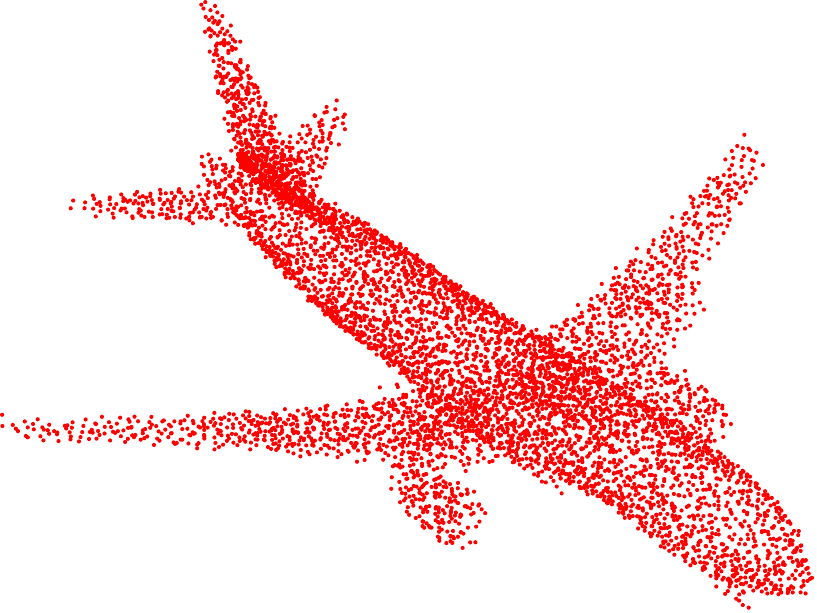} &
		\includegraphics[width=\figw]{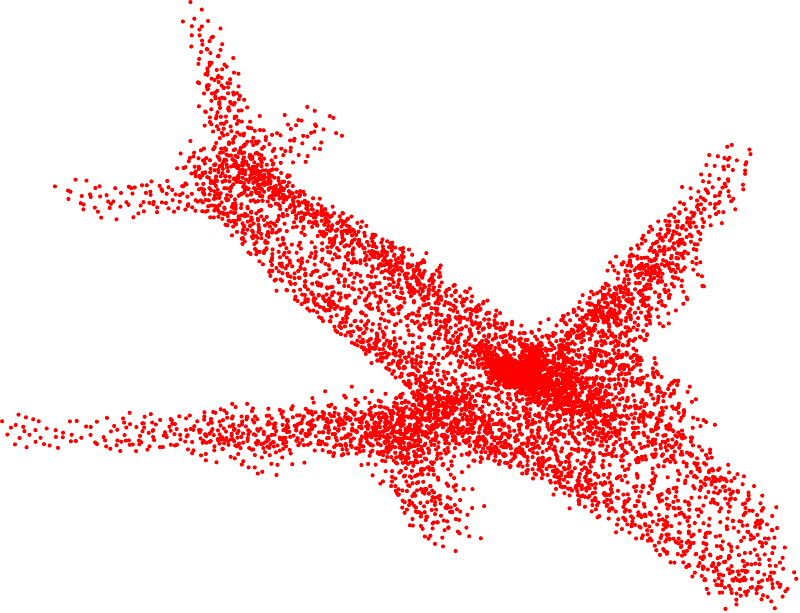} &
		\includegraphics[width=\figw]{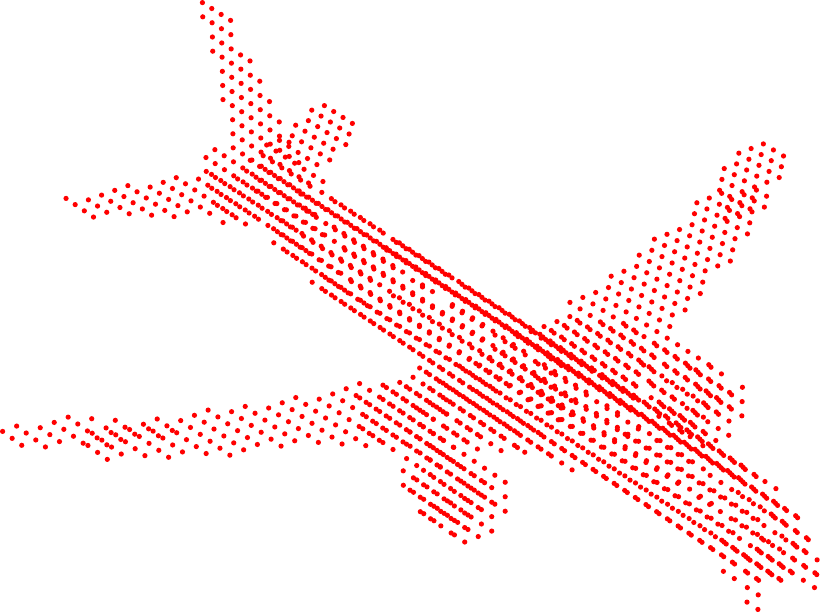} &
		\includegraphics[width=\figw]{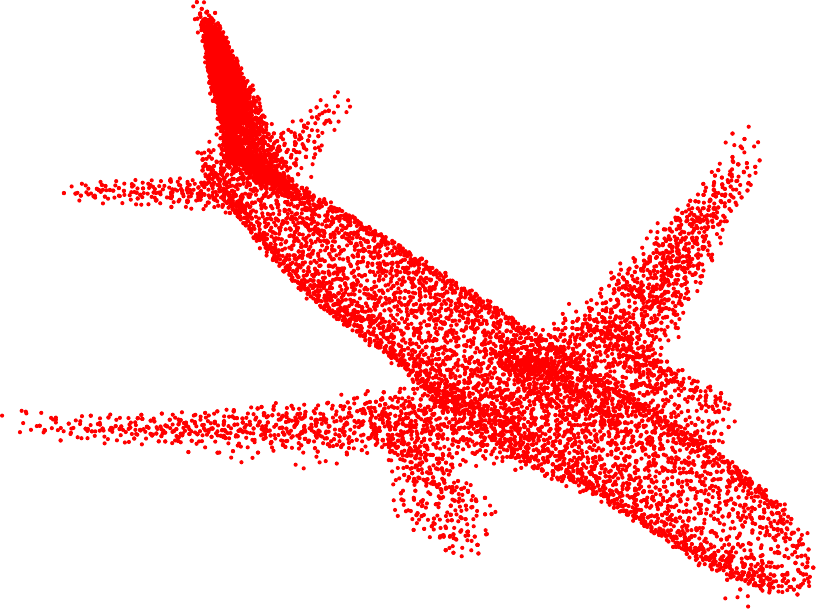} \\[-1.5pt]
		
		\makecell{\scriptsize airplane} &
		\makecell{
			\scriptsize D1 43.7777 \\ 
			\scriptsize D2 46.9665 \\ 
			\scriptsize CD 1.12e-4} &
		\makecell{
			\scriptsize D1 33.2278 \\ 
			\scriptsize D2 35.1125 \\ 
			\scriptsize CD 13.45e-4} &
		\makecell{
			\scriptsize D1 38.5904 \\[-1.5pt]
			\scriptsize D2 39.6715 \\[-1.5pt]
			\scriptsize CD 3.97e-4} &
		\makecell{
			\scriptsize D1 44.0464 \\ 
			\scriptsize D2 47.3436 \\ 
			\scriptsize CD 1.09e-4} \\
		
		% ================== 2. bathtub ==================
		\includegraphics[width=\figw]{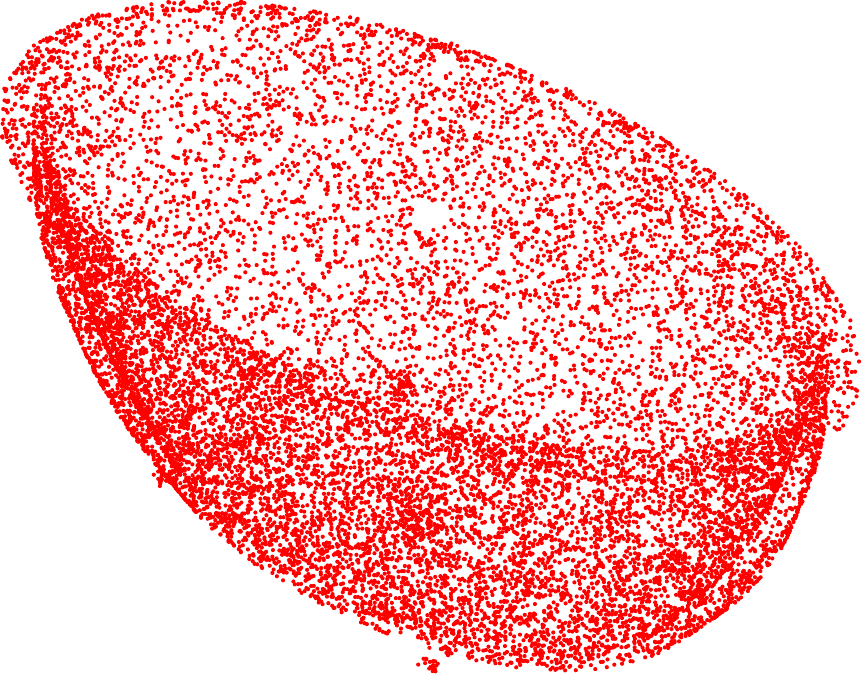} &
		\includegraphics[width=\figw]{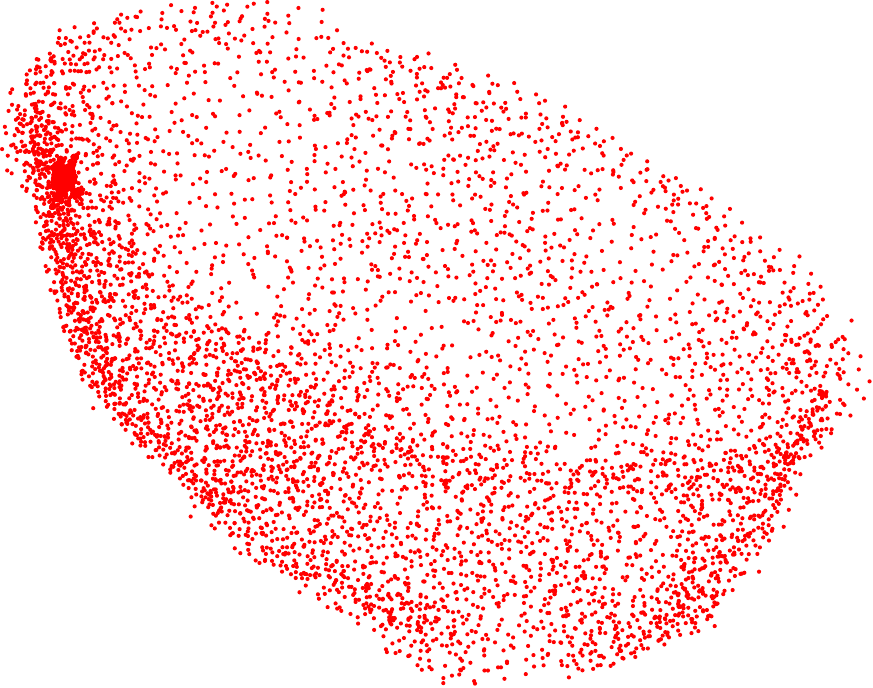} &
		\includegraphics[width=\figw]{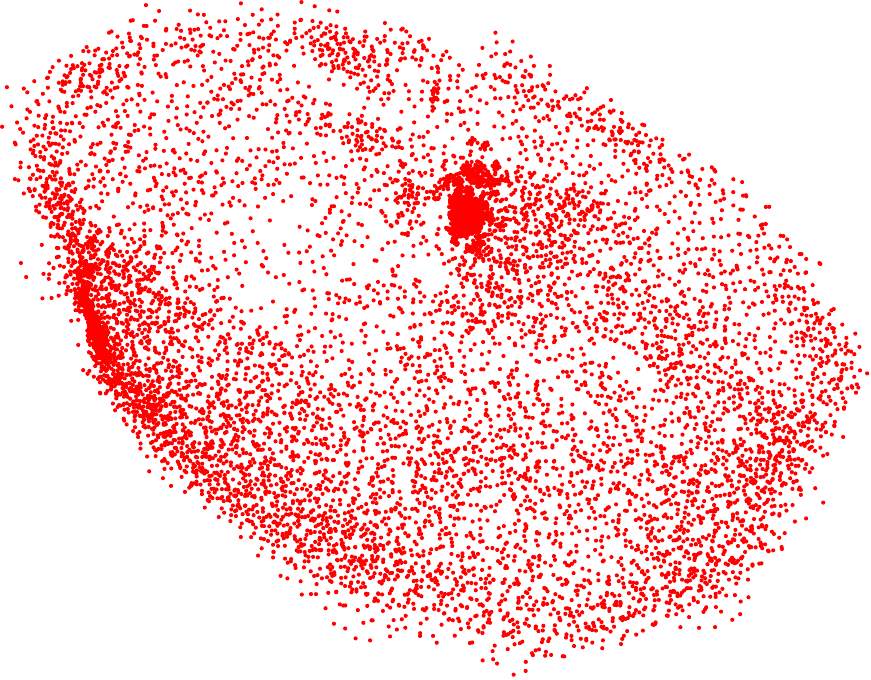} &
		\includegraphics[width=\figw]{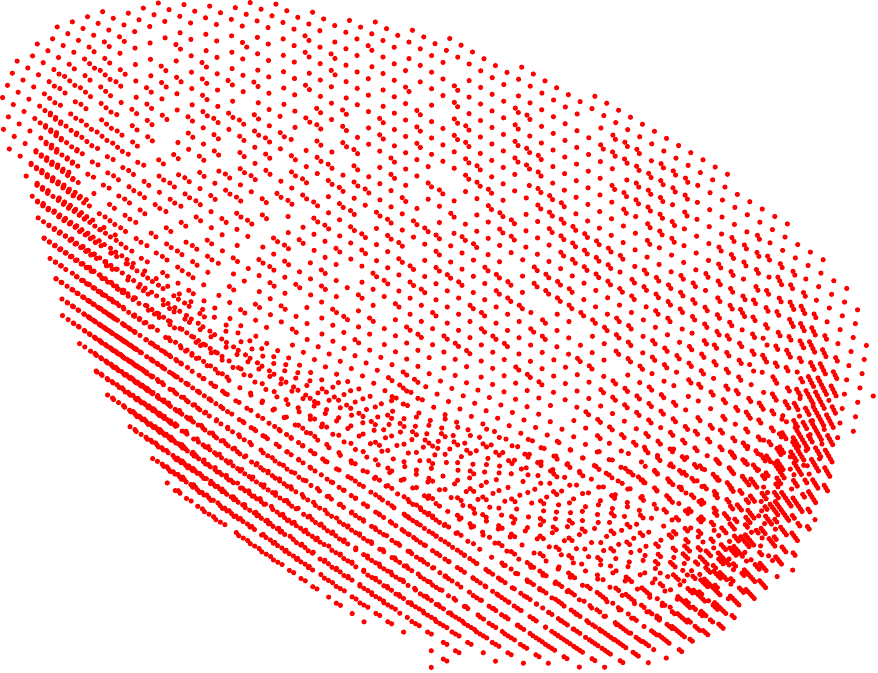} &
		\includegraphics[width=\figw]{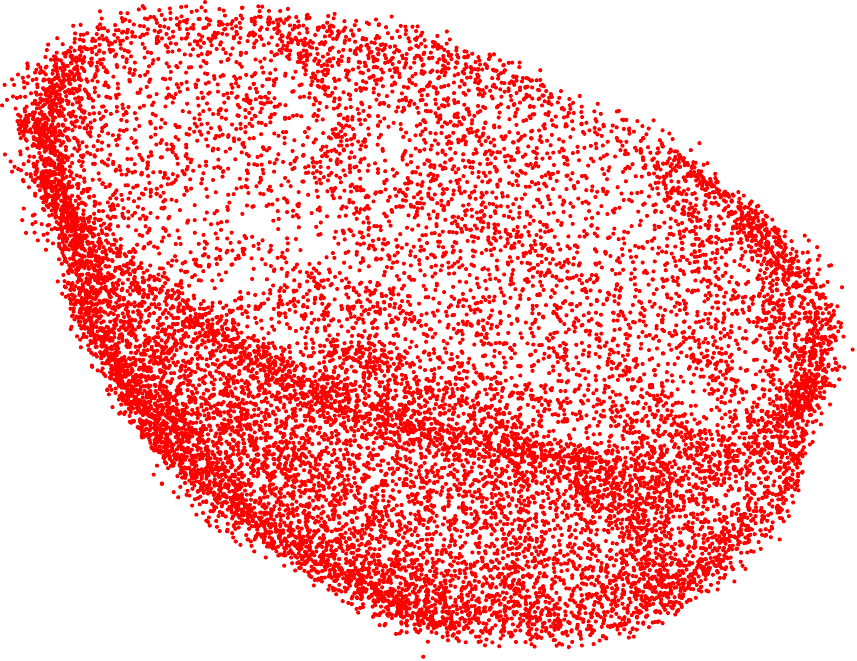} \\ 
		
		\makecell{\scriptsize bathtub} &
		\makecell{\scriptsize D1 37.9470 \\  \scriptsize D2 41.3004 \\  \scriptsize CD 4.34e-4} &
		\makecell{\scriptsize D1 33.8456 \\  \scriptsize D2 34.4114 \\  \scriptsize CD 9.43e-4} &
		\makecell{\scriptsize D1 37.0805 \\  \scriptsize D2 40.4201 \\  \scriptsize CD 4.36e-4} &
		\makecell{\scriptsize D1 39.5759 \\  \scriptsize D2 42.7453 \\  \scriptsize CD 2.71e-4} \\
		
		% ================== 3. chair ==================
		\includegraphics[width=0.16\linewidth]{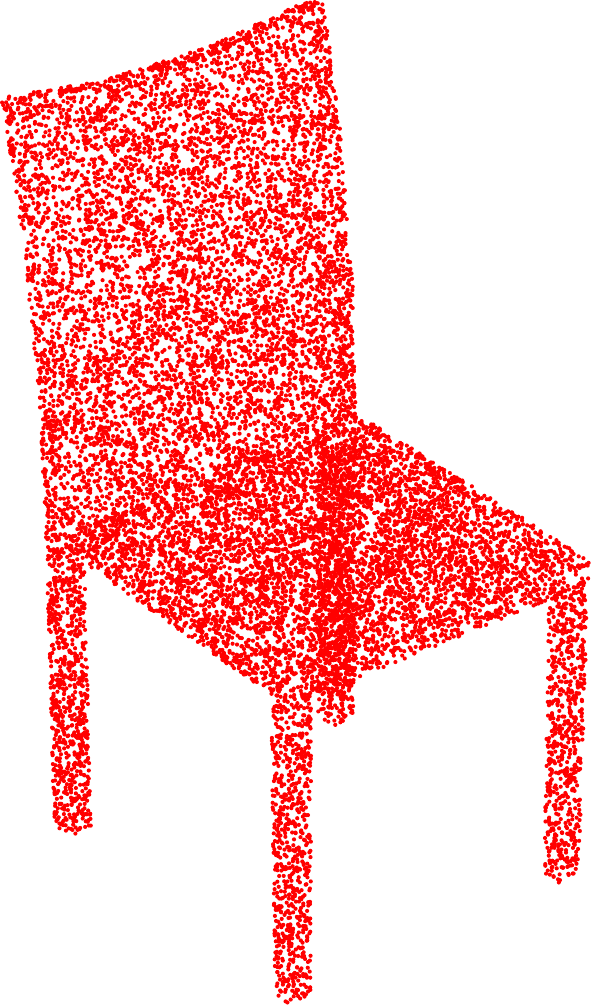} &
		\includegraphics[width=0.16\linewidth]{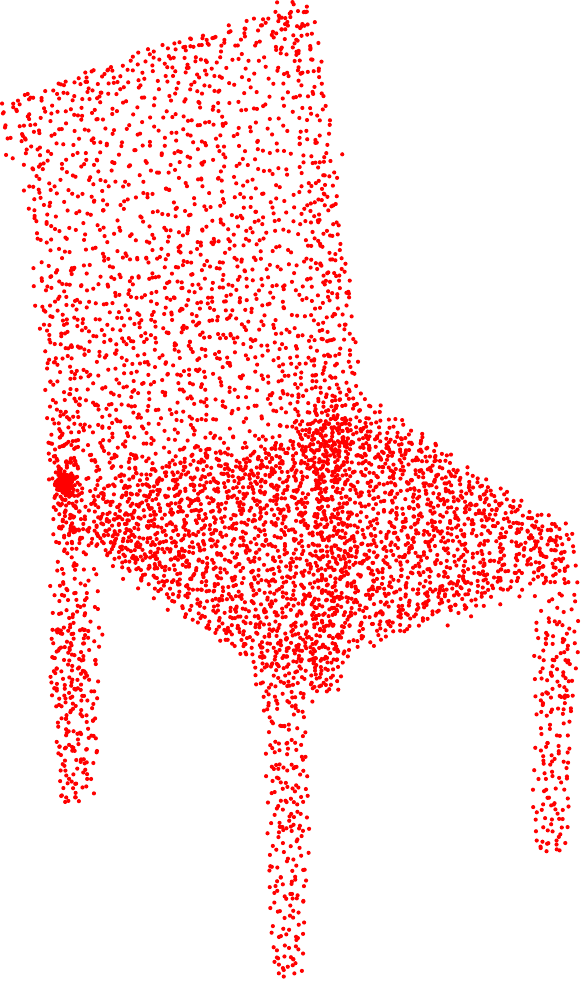} &
		\includegraphics[width=0.16\linewidth]{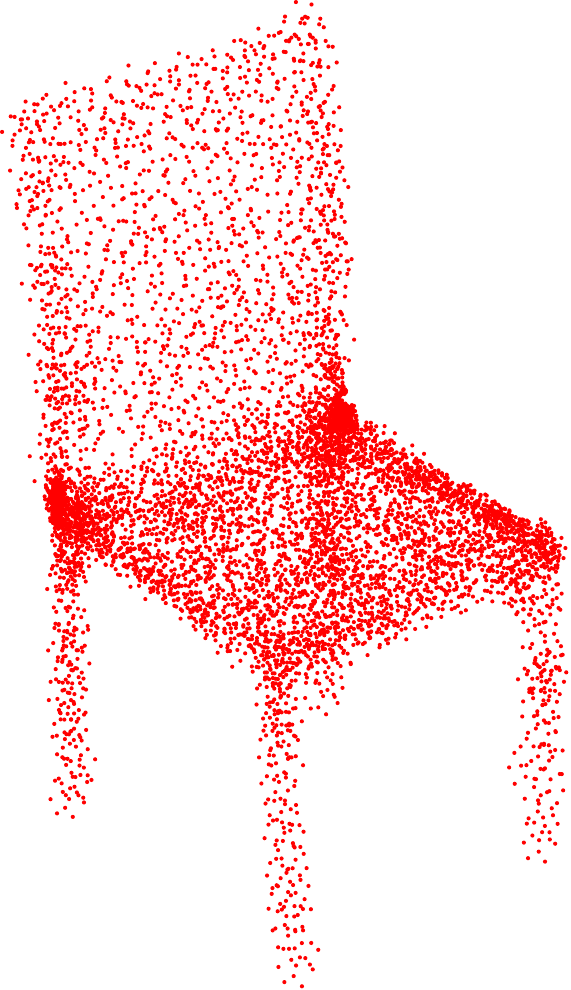} &
		\includegraphics[width=0.16\linewidth]{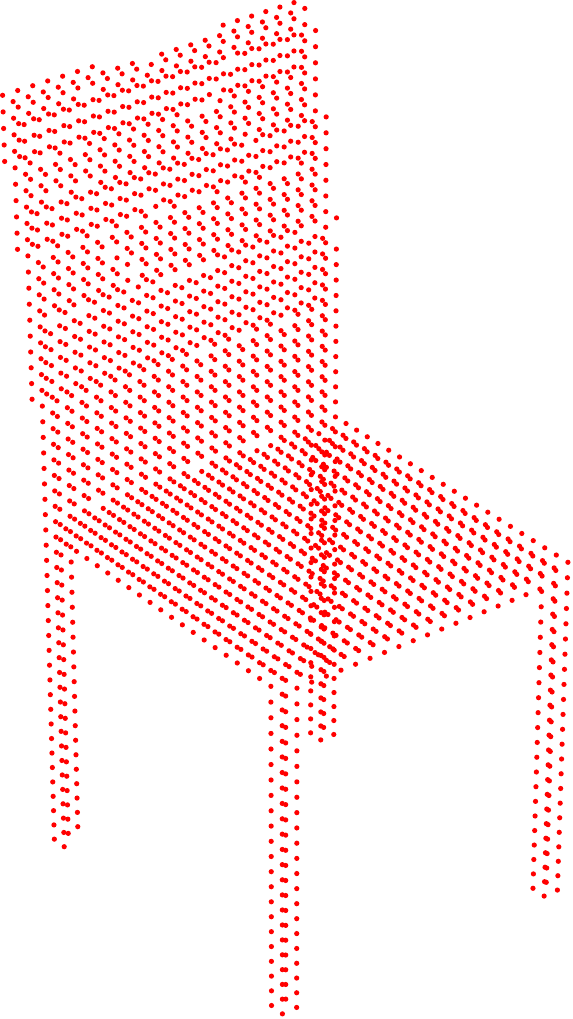} &
		\includegraphics[width=0.16\linewidth]{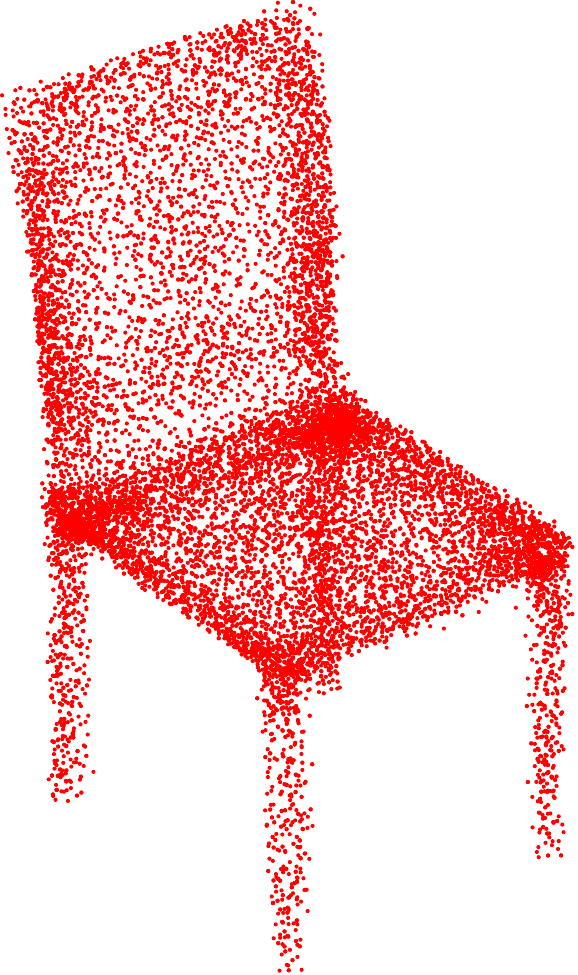} \\ 
		
		\makecell{\scriptsize chair} &
		\makecell{\scriptsize D1 39.4404 \\  \scriptsize D2 42.6084 \\  \scriptsize CD 2.52e-4} &
		\makecell{\scriptsize D1 34.3449 \\  \scriptsize D2 35.2081 \\  \scriptsize CD 9.63e-4} &
		\makecell{\scriptsize D1 37.5876 \\  \scriptsize D2 40.7382 \\  \scriptsize CD 4.16e-4} &
		\makecell{\scriptsize D1 40.4505 \\  \scriptsize D2 43.3372 \\  \scriptsize CD 2.32e-4} \\
		
		% ================== 4. toilet ==================
		\includegraphics[width=\figw]{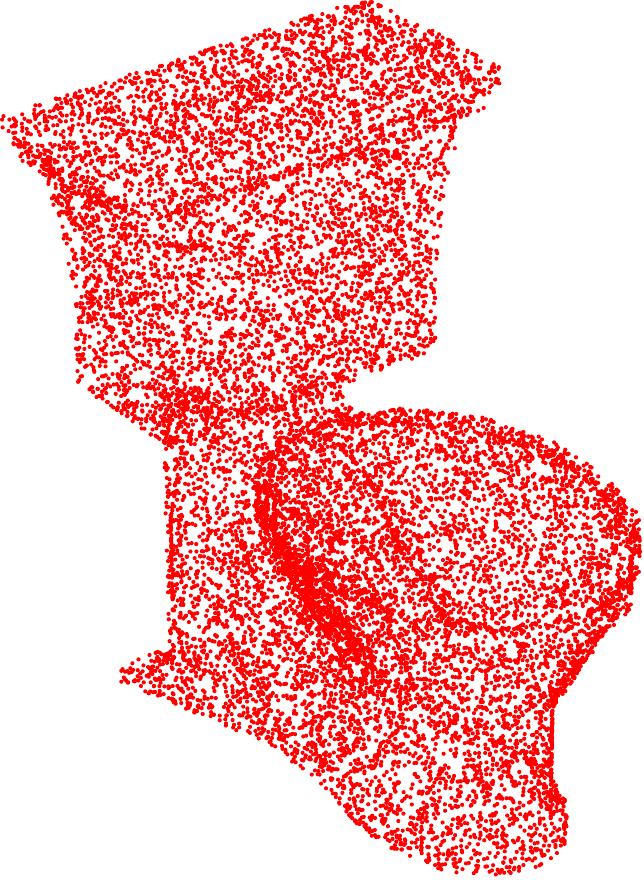} &
		\includegraphics[width=\figw]{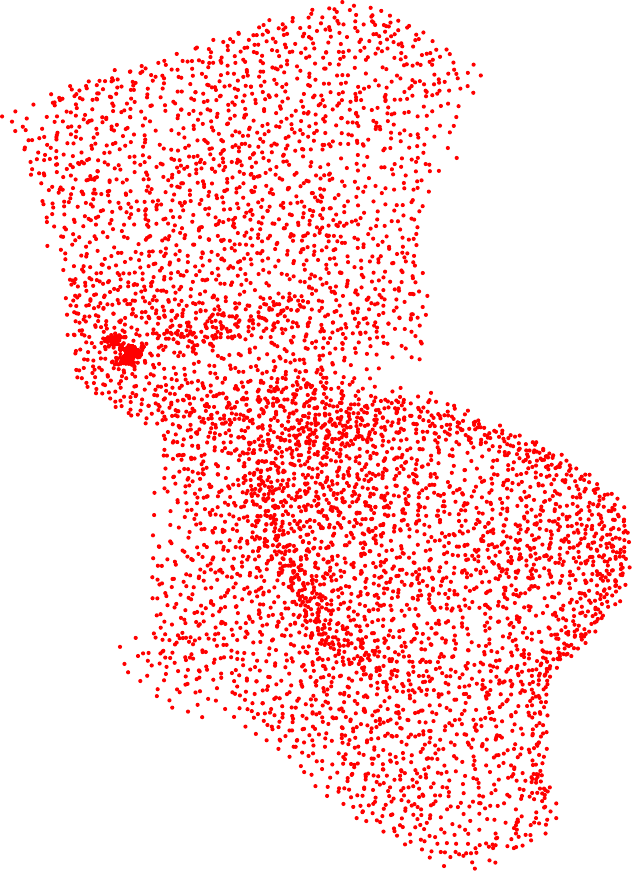} &
		\includegraphics[width=\figw]{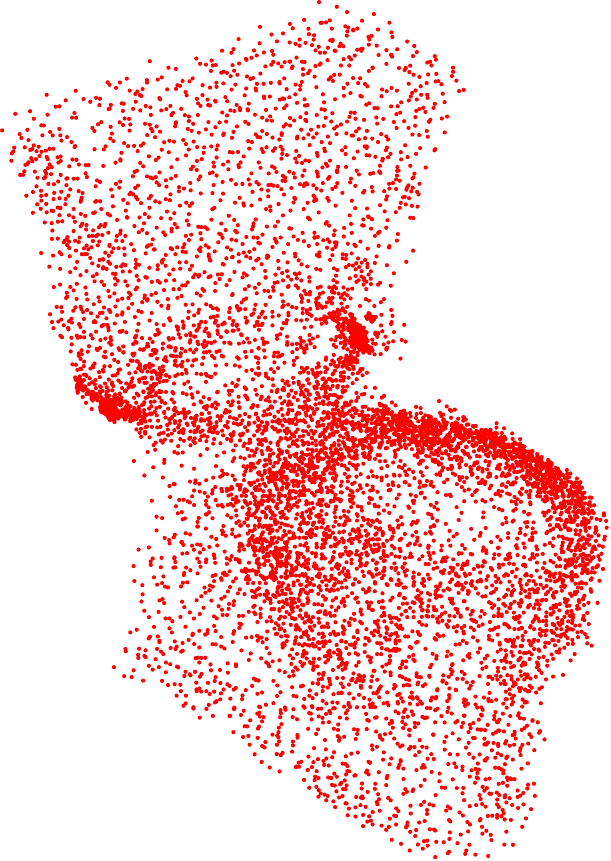} &
		\includegraphics[width=\figw]{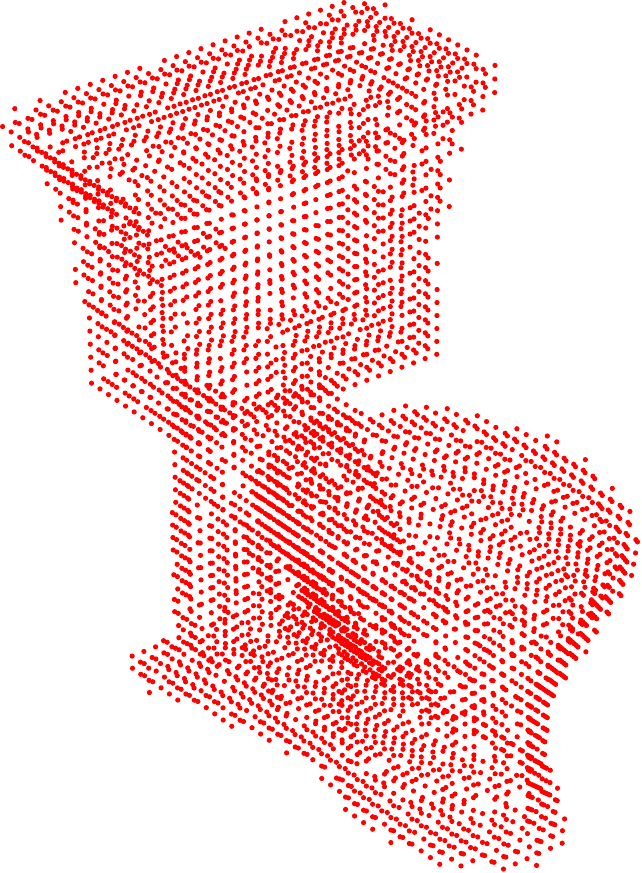} &
		\includegraphics[width=\figw]{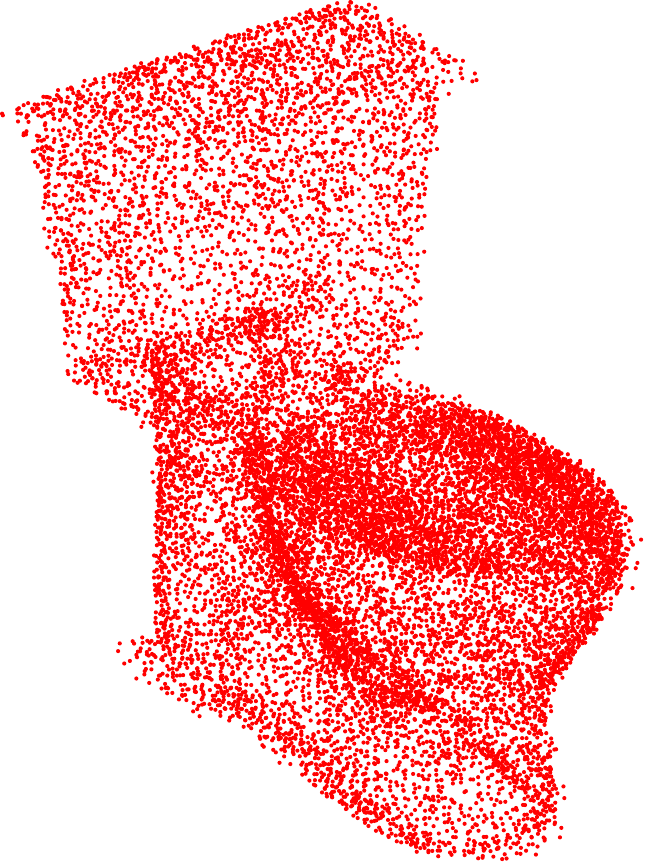} \\ 
		
		\makecell{\scriptsize toilet} &
		\makecell{\scriptsize D1 36.6270 \\  \scriptsize D2 40.3114 \\  \scriptsize CD 4.79e-4} &
		\makecell{\scriptsize D1 33.8806 \\  \scriptsize D2 37.4858 \\  \scriptsize CD 10.25e-4} &
		\makecell{\scriptsize D1 35.2298 \\  \scriptsize D2 39.9090 \\  \scriptsize CD 5.39e-4} &
		\makecell{\scriptsize D1 37.3230 \\  \scriptsize D2 40.5825 \\  \scriptsize CD 4.71e-4} \\
			
	\end{tabular}
	\caption{Visual and quantitative comparisons of reconstructed point clouds at SNR $=10$ dB and $n=512$. G-PCC is shown only as a relaxed-budget operability reference with $n=5000$.
	Corresponding D1-PSNR, D2-PSNR, and CD values are reported below each example.
	}
	\label{fig:benchmark_visual}
\end{figure}

The qualitative comparisons with different baselines are shown in Fig. \ref{fig:benchmark_visual}, obtained at SNR $=10$ dB.
The visual results are consistent with the quantitative trends.
SEPT generally preserves the coarse silhouette of each object, but it produces uneven point density.
DPCC is more vulnerable to outliers and geometric distortion, indicating that its representation is less stable after MIMO transmission.
The G-PCC reference remains visibly sparse even under the much larger bandwidth budget.
By comparison, POST-MIMO preserves part connectivity, surface smoothness, and fine local details more faithfully.
These observations further support that the proposed block-aligned design makes more effective use of the available bandwidth for geometry preservation.

Table \ref{tab:performance} reports the computational cost of the learned schemes at $n=512$, including inference time (Time) in milliseconds (ms), trainable parameters (Params.) in millions (M), floating point operations (FLOPs) in billions (B), and the peak GPU memory in megabytes (MB).
\begin{table}[!t]
	\centering
	\caption{Complexity and efficiency comparison of different schemes.}
	\label{tab:performance} % 总标签

	\begin{tabular}{|l|c|c|c|}
		\hline
		Schemes & SEPT & DPCC & POST-MIMO \\
		\hline
		Time (ms) & 65.92 & 3208.65 & 24.96 \\
		\hline
		Params. (M) & 274.24 & 1077.57 & 35.24 \\
		\hline
		FLOPs (B) & 18.42 & 110.41 & 18.09 \\
		\hline
		\makecell{Peak GPU Memory (MB)} & 7046 & 8752 & 2484 \\
		\hline
	\end{tabular}
\end{table}
The reported results are based on our unified re-implementations and MIMO adaptations; G-PCC is omitted due to its substantially larger bandwidth requirement, while the DPCC results exclude digital modulation overhead.
POST-MIMO achieves the lowest latency, parameter count, and memory usage with FLOPs comparable to SEPT by performing expensive operations on compact coarse representations and using lightweight modules for progressive dense reconstruction.

\begin{figure}[!t]
	\centering
	\subfloat[]{
		\includegraphics[width=0.3\textwidth]{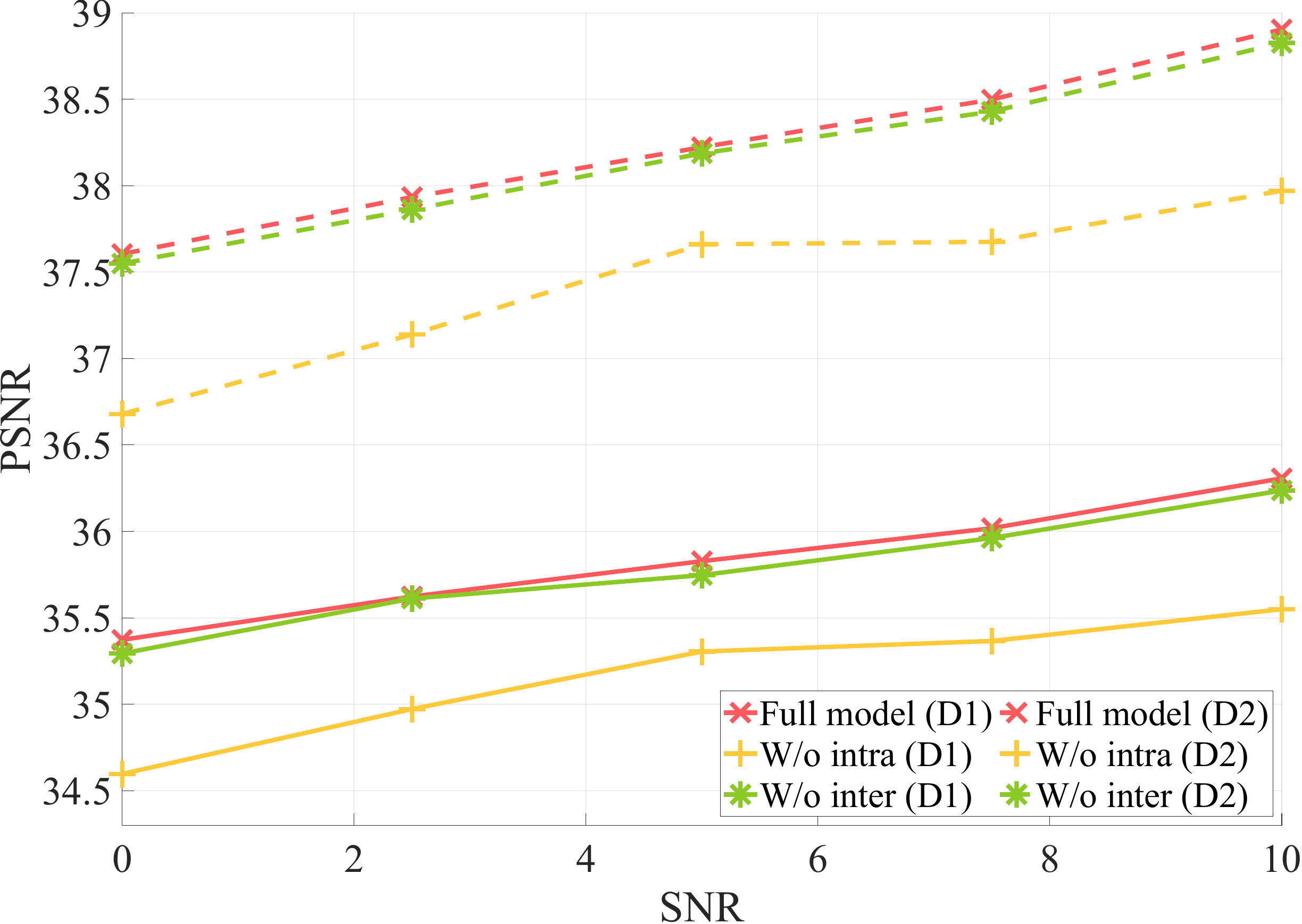}%
		\label{subfig:ablation1}
	}
	\hfil
	\subfloat[]{
		\includegraphics[width=0.3\textwidth]{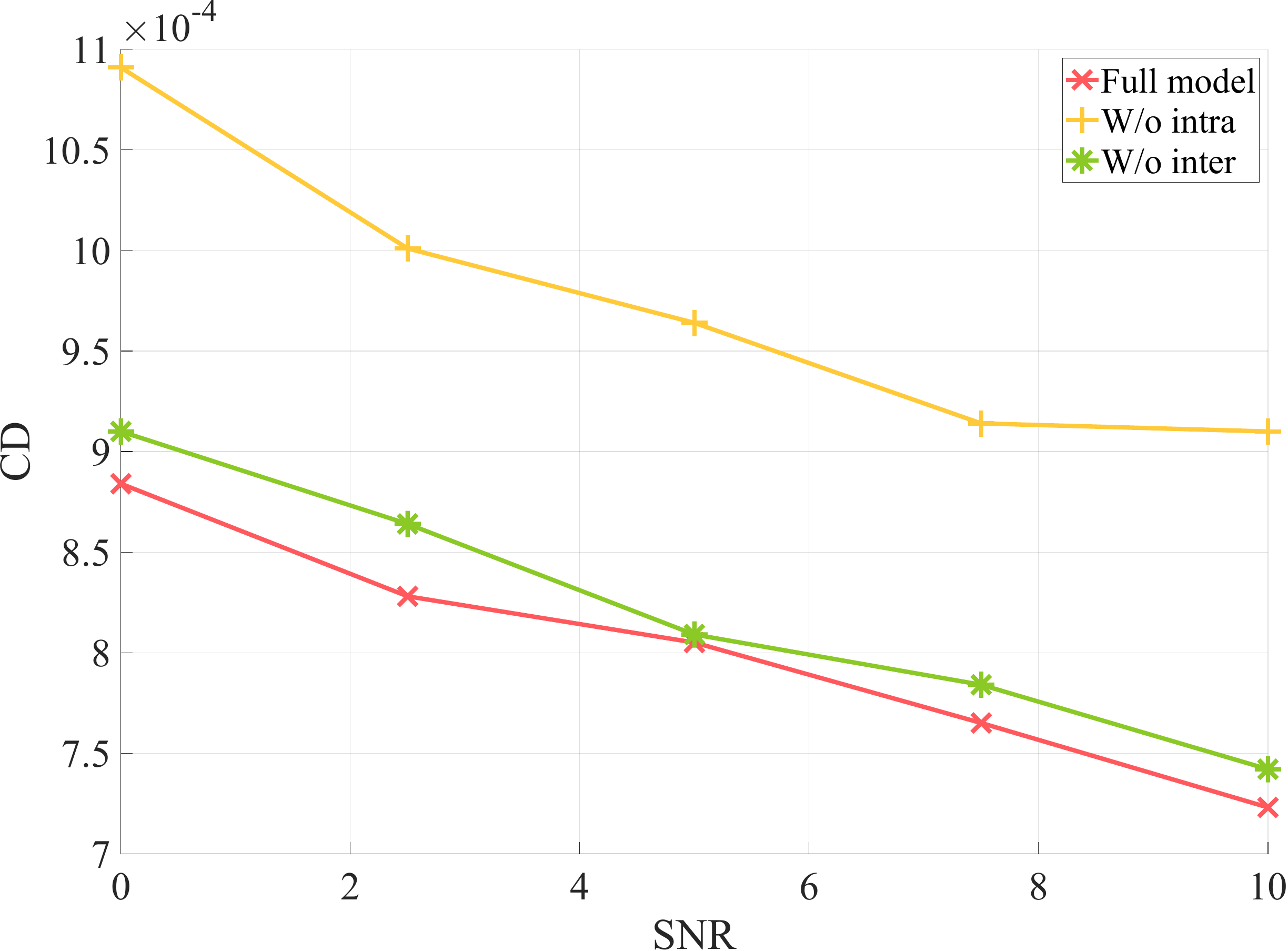}%
		\label{subfig:ablation2}
	}
	\caption{Ablation study of the proposed framework under varying SNR.
	(a) D1/D2.
	(b) CD.}
	\label{fig:ablation}
\end{figure}

We further evaluate the contribution of the encoder, including the intra-block local extractor (W/o intra) and the inter-block context module (W/o inter).
As shown in Fig.~\ref{fig:ablation}, both modules are necessary for effective transmission.
The performance drop is larger without the intra-block module, indicating that local geometry extraction is more fundamental to the learned representation, while the degradation without inter-block interaction confirms its role in preserving global structural coherence.

\section{Conclusion}
This paper presents POST-MIMO for semantic point cloud transmission over MIMO channels.
By aligning spatial blocks with parallel streams, POST-MIMO combines local and global feature extraction at the transmitter with progressive geometry reconstruction at the receiver.
The experimental results show that this structured design yields gains in reconstruction quality over representative baselines.
Under a representative setting of SNR $=10$ dB and $n=512$, POST-MIMO reduces CD by at least 32.11\% relative to the baselines, while also substantially lowering computational cost.
Overall, the proposed framework provides a promising design for wireless point cloud transmission over MIMO channels.

{
    \small
    \bibliographystyle{IEEEtran}{\tiny}
    \bibliography{IEEEabrv, main_ref}
}

\end{document}